\documentclass[11pt]{article}
\usepackage{amsmath,amssymb,amsthm,empheq}
\usepackage[dvipdfmx]{graphicx}
\usepackage{comment}
\usepackage{fancybox}
\usepackage{eurosym}
\usepackage{type1cm}
\usepackage{url}

\renewcommand{\Vec}[1]{\mbox{\boldmath $#1$}}
\newcommand{\Real}{\mathbb{R}}
\newcommand{\Int}{\mathbb{Z}}
\newcommand{\deT}{\mbox{\rm det}}
\newcommand{\minW}{{\cal W}^{\mbox{m}}}
\newcommand{\maxL}{{\cal L}^{\mbox{M}}}
\newcommand{\overQ}{$\overline{\mbox{\rm Q}}$}
\newcommand{\QED}{\hfill QED}
\newcommand{\Exp}{\mbox{\rm E}}
\newcommand{\dd}{\mbox{\rm d}}

\newcommand{\citep}[1]{\cite{#1}}

\newtheorem{theorem}{Theorem}[section]

\newtheorem{lemma}[theorem]{Lemma}

\title{Trading Transforms 
	of Non-weighted Simple Games \\
	and Integer Weights 
	of Weighted Simple Games
\thanks{
 preliminary version of this paper was presented at 
Seventh International Workshop on Computational Social Choice (COMSOC-2018), Rensselaer Polytechnic Institute, Troy, NY, USA, 25-27 June, 2018.  
}
}

\author{Akihiro Kawana 
\thanks{Graduate School of Engineering, 
   Tokyo Institute of Technology 
}
       \and
        Tomomi Matsui 
\thanks{Graduate School of Engineering, 
   Tokyo Institute of Technology 
}
}

\begin{document}
\maketitle

\begin{abstract}
This study investigates simple games. 
A fundamental research question in this field is 
	to determine necessary and sufficient conditions for
	a simple game to be a weighted majority game. 
Taylor and Zwicker (1992) 
	showed that a simple game is non-weighted 
	if and only if there exists a trading transform 
	of finite size. 
They also provided an upper bound 
	on the size of such a trading transform, if it exists. 
Gvozdeva and Slinko (2011) improved that upper bound; 
	their proof employed a property of linear inequalities 
	demonstrated by Muroga (1971).
In this study, we provide a new proof of the existence 
	of a trading transform 
	when a given simple game is non-weighted. 
Our proof employs Farkas' lemma (1894), 
	and yields an improved upper bound 
	on the size of a trading transform.

We also discuss an integer-weight representation
	of a weighted simple game, improving the bounds 
	obtained by Muroga (1971).
We show that our bound on the quota is tight when the number 
	of players is less than or equal to five,
	based on the computational results obtained by Kurz (2012).

Furthermore, we discuss the problem of finding
	an integer-weight representation 
	under the assumption that
	we have minimal winning coalitions 
	and maximal losing coalitions.
In particular, we show a performance of a rounding method.

Lastly, we address roughly weighted simple games.
Gvozdeva and Slinko (2011) showed that 
	a given simple game is not roughly weighted 
	if and only if 
	there exists a potent certificate of non-weightedness.
We give an upper bound on 	the length of 
	a potent certificate of non-weightedness.
We also discuss an integer-weight representation
	of a roughly weighted simple game.

\end{abstract}

\section{Introduction}

A simple game consists of a pair $G=(N, {\cal W}),$
	where $N$ is a finite set of players, and 
	${\cal W} \subseteq 2^N$ is an arbitrary collection
	of subsets of $N$.
Throughout this paper, we denote $|N|$ by $n$.
Usually, the  property
\begin{equation}\label{monotonicity}
\mbox{(monotonicity):}  
	\mbox{ if } S' \supseteq S \in {\cal W},
	\mbox{ then } S' \in {\cal W},
\end{equation}
is assumed.
\noindent 
Subsets in ${\cal W}$ are called {\em winning coalitions}.
We denote $2^N \setminus {\cal W}$ by ${\cal L}$, 
	and subsets in ${\cal L}$ are called
	{\em losing coalitions}.
A simple game $(N, {\cal W})$ is said to be 
	{\em weighted} if there exists 
	a weight vector $\Vec{w} \in  \Real^N$ and $q \in \Real$
	satisfying the following property: 
\begin{equation}\label{def-weightedness}
\mbox{(weightedness):}  
\mbox{ for any } S \subseteq N, 
		S \in {\cal W} \mbox{ if and only if } 
		\sum_{i \in S} w_i \geq q.
\end{equation}

Previous research established the necessary and sufficient conditions 
	that guarantee the weightedness of a simple.
\cite{elgot1961decision}
	and ~\cite{chow1961characterization} 
	investigated the theory of threshold logic
	and showed the condition of the weightedness  
	in terms of {\em asummability}. 
\cite{muroga1971threshold} proved the sufficiency of asummability
	based on the theory of linear inequality systems
	and discussed some variations of their results
	in cases of a few variables.
\cite{taylor1992characterization,taylor1999simple} 
	obtained 
	necessary and sufficient conditions independently 
	in terms of a {\em trading transform}. 
A {\em trading transform} of size $j$
	 is a coalition sequence 
	$(X_1, X_2, \ldots, X_j; Y_1, Y_2,\ldots, Y_j)$,
	which may contain repetitions of coalitions,
	satisfying the condition $\forall p \in N$, 
	$|\{i \mid p \in X_i\}|=|\{i \mid p \in Y_i\}|$.
A simple game is called {\em $k$-trade robust} 
	if there is no trading transform of size $j$ 
	satisfying  $1 \leq j \leq k$, 
	 $X_1, X_2, \ldots , X_j \in {\cal W}$, and
	 $Y_1, Y_2, \ldots , Y_j \in {\cal L}$. 
A simple game is called {\em trade robust}
	 if it is $k$-trade robust for all positive integers  $k$.

Taylor and Zwicker showed that
	a given simple game $G$ with $n$ players 
	is weighted
	if and only if 
	$G$ is $2^{2^n}$-trade robust.
In 2011, \cite{gvozdeva2011weighted}
	showed that 
	a given simple game $G$ is weighted
	if and only if 
	$G$ is $(n+1)n^{n/2}$-trade robust.
\cite{freixas2009simple}	 
	proposed a variant of trade robustness, 
	called invariant-trade robustness, 
	which determines 
	whether a simple game is weighted.
The relations between the results 
	in threshold logic
	and simple games are clarified
	in~\cite{freixas2016characterization,freixas2017characterization}. 

In Section~\ref{tradingtransform},  
	we show that a given simple game $G$ is weighted
	if and only if 
	$G$ is $\alpha_{n+1}$-trade robust, 
	where $\alpha_{n+1}$ denotes 
	the maximal value of determinants of 
	$(n+1) \times (n+1)$ 0--1 matrices.
It is well-known that
	$\alpha_{n+1} \leq (n+2)^{\frac{n+2}{2}}(1/2)^{(n+1)}$.

Our definition of a weighted simple game allows
	for an arbitrary real number of weights.
However,  any weighted simple game 
	can be represented by integer weights
	(e.g., see \cite{freixas2009existence}).
An {\em integer-weight representation}
	of a weighted simple game
	consists of an integer vector 
	$\Vec{w} \in  \Int^N$ and some $q \in \Int$
	satisfying the weightedness property~(\ref{def-weightedness}).
\cite{isbell1956class} found an example 
	of a weighted simple game 
	with 12 players without a unique minimum-sum 
	integer-weight representation. 
Examples for 9, 10, or 11 players are given 
	in~\cite{freixas2009existence,freixas2010weighted}.
In the field of threshold logic, 
	examples of threshold functions
	requiring large weights are discussed
	by~\cite{myhill1961size,muroga1971threshold,haastad1994size}.
Some previous studies enumerate 
	(minimal) integer-weight representations
	of simple games with a small number of players 
	(e.g.,~\cite{muroga1962majority,winder1965enumeration,%
muroga1970enumeration,krohn1995directed}).
In the case of $n=9$ players, 
	refer to~\cite{kurz2012minimum}.
In general, \cite{muroga1971threshold} 
	(Proof of Theorem 9.3.2.1) showed that
	(under the monotonicity property~(\ref{monotonicity}) and 
	$\emptyset \not \in {\cal W} \ni N$)
	every weighted simple game has 
	an integer-weight representation 
	satisfying 
	 $0\leq w_i \leq \alpha_n \leq (n+1)^{\frac{n+1}{2}}(1/2)^n$ 
	 	$(\forall i \in N)$
		and
	$0 \leq q \leq n \alpha_n  \leq n(n+1)^{\frac{n+1}{2}}(1/2)^n$ 
	simultaneously. 
Here,  $\alpha_n$ denotes 
	the maximal value of determinants of 
	$n \times n$ 0--1 matrices.
\cite{wang1991threshold}
	discussed Boolean functions that require more general
	surfaces to separate their true vectors 
	from false vectors.
\cite{hansen2015polynomial}  investigates the complexity 
	of computing Boolean functions by polynomial threshold functions. 
\cite{freixas2020characterization} discusses 
a point-set-additive pseudo-weighting for a simple game, 
which assigns weights directly to coalitions.

In Section~\ref{integerweights},
	we slightly improve Muroga's result 
	and 
	show that every weighted simple game 
	(satisfying $\emptyset \not \in {\cal W} \ni N$)
	has an integer-weight representation 
	$(q; \Vec{w}^{\top})$ 
	satisfying 
	$|w_i| \leq \alpha_n$ $(\forall i \in N)$, 
	$|q| \leq \alpha_{n+1}$, 
	and 
	$1 \leq \sum_{i \in N} w_i \leq 2 \alpha_{n+1}-1$
	simultaneously. 
Based on the computational results 
	of~\cite{kurz2012minimum},
	we also demonstrate the tightness of our bound on the quota 
	when $n \leq 5$.

For a family of minimal winning coalitions,
	\cite{peled1985polynomial}
	proposed a polynomial-time algorithm
	for checking the weightedness of a given simple game.
They also showed that for weighted simple games
	represented by minimal winning coalitions, 
	all maximal losing coalitions can be computed 
	in polynomial time.
When we have minimal winning coalitions 
	and maximal losing coalitions,
	there exists a linear inequality system whose solution gives  
	a weight vector $\Vec{w} \in  \Real^N$ and $q \in \Real$
	satisfying property~(\ref{def-weightedness}).
However, it is less straightforward to find 
	an integer-weight representation 
	as the problem transforms 
	from linear programming to integer programming.

In Section~\ref{approximation},
	we address the problem of finding
	an integer-weight representation 
	under the assumption that
	we have minimal winning coalitions 
	and maximal losing coalitions.
We show that
	an integer-weight representation is obtained 
	by carefully rounding a solution of the linear inequality system
	multiplied by at most $ (2-\sqrt{2})n+(\sqrt{2}-1)$.
	

A simple game 	$G=(N, {\cal W})$ is called 
	{\em roughly weighted} 
	if there exist a non-negative vector 
	$\Vec{w}\in \Real_+^N$ and a real number $q \in \Real$, 
	not all equal to zero ($(q; \Vec{w}^{\top})\neq \Vec{0}^{\top}$), 
	such that 
	for any $S \subseteq N$ 
	condition $\sum_{i \in S}  w_i < q$ implies $S \not \in {\cal W}$,
	and $\sum_{i \in S} w_i > q$ implies $S \in {\cal W}$. 
We say that $(q; \Vec{w}^{\top})$  is a
	{\em rough voting representation} for $G$.
Roughly weighted simple games were initially introduced 
	by~\cite{baugh1970pseudo}.
\cite{muroga1971threshold} (p. 208)  studied them 
	under the name of pseudothreshold functions.
\cite{taylor1999simple}  discussed roughly weighted simple games 
	 and constructed several examples.
\cite{gvozdeva2011weighted} developed a theory 
	of roughly weighted simple games.	
A trading transform $(X_1, X_2, \ldots, X_j; Y_1, Y_2,\ldots, Y_j)$ 
	with all coalitions $X_1, X_2, \ldots , X_j$ winning and  
	$Y_1, Y_2, \ldots , Y_j$ losing 
	is called a {\em certificate of non-weightedness}.
This certificate is said to be {\em potent} 
	if the grand coalition $N$ is among  $X_1, X_2, \ldots , X_j$ 
	and the empty coalition is among $Y_1, Y_2, \ldots , Y_j$.
\cite{gvozdeva2011weighted} showed that  
	under the the monotonicity property~(\ref{monotonicity}) and
	$\emptyset \not \in {\cal W} \ni N$, 
	a given simple game $G$ is not roughly weighted 
	if and only if 
	there exists a potent certificate of non-weightedness 
	whose length is less than or equal to $(n+1)n^{n/2}$. 
Further research on roughly weighted simple games 
	appears in 	
	\cite{gvozdeva2013three,freixas2014alpha,hameed2015roughly}.

In Section~\ref{RWSG}, we show that 
	(under the the monotonicity
	 property~(\ref{monotonicity})	 and $\emptyset \not \in {\cal W} \ni N$) 
	the length of a potent certificate of non-weightedness 
	is less than or equal to 
	$ 2 \alpha_{n+1}$, if it exists. 
We also show that
	a roughly weighted simple game 
	(satisfying $\emptyset \not \in {\cal W} \ni N$)
	has an integer vector $(q; \Vec{w}^{\top})$ 
	of rough voting representation  satisfying
	$0 \leq w_i \leq \alpha_{n-1}$ 
	$(\forall i \in N)$,  
	$0 \leq q \leq \alpha_{n}$ 
	and $0 \leq \sum_{i \in N} w_i \leq 2 \alpha_{n}$.


\section{Trading Transforms of Non-weighted Simple Games}\label{tradingtransform}

In this section, we discuss the size of a trading transform
	that guarantees the non-weightedness of a given simple game.
Throughout this section, we do not need to assume 
	the monotonicity property~(\ref{monotonicity}).	
First, we introduce a linear inequality system
	for determining the weightedness of a given simple game.
For any nonempty family of player subsets 
	$\emptyset \neq {\cal N} \subseteq 2^N$,
	we introduce a 0--1 matrix $A({\cal N})=(a({\cal N})_{Si})$
	whose rows are indexed by subsets in ${\cal N}$ 
	and columns are indexed by players in $N$ defined by
\[
		a({\cal N})_{Si}=
		\left\{
		\begin{array}{ll}
			1 & (\mbox{if } i \in S \in {\cal N}), \\
			0 & (\mbox{otherwise}).
		\end{array}
		\right.
\]
\noindent
A given simple game $G=(N,{\cal W})$
	is weighted if and only if 
	the following linear inequality system is feasible: 
\begin{eqnarray*}
	\mbox{P1:}
	\left(
	\begin{array}{rrr}
		 A({\cal W})&  \Vec{1} &  \Vec{0} \\
		-A({\cal L})& -\Vec{1} & -\Vec{1} 
	\end{array}
	\right)
	\left(
	\begin{array}{r}
		\Vec{w} \\ - q \\ \varepsilon 
	\end{array}
	\right)
	&\geq & \Vec{0}, \\
	\varepsilon & > & 0,
\end{eqnarray*}
	where $\Vec{0}$  ($\Vec{1}$) denotes
	a zero vector (all-one vector) 
	of an appropriate dimension.

Farkas' Lemma~\citep{farkas1902theorie} states that
	P1 is infeasible if and only if the following system is feasible:
\begin{eqnarray*}
	\mbox{D1:}
	\left(
	\begin{array}{cc}
		A({\cal W})^{\top} & -A({\cal L})^{\top} \\
		 \Vec{1}^{\top} 		& -\Vec{1}^{\top} \\
	   \Vec{0}^{\top} 		& -\Vec{1}^{\top}
	\end{array}
	\right)
	\left(
	\begin{array}{c}
		\Vec{x}  \\ \Vec{y}
	\end{array}
	\right)
	& = & 
	\left(
	\begin{array}{c}
		\Vec{0} \\ 0 \\ -1
	\end{array}
	\right), \\
	\Vec{x} \geq \Vec{0}, \;\; \Vec{y} \geq \Vec{0}.
\end{eqnarray*} 
For simplicity, 
	we denote D1 by 
	$A_1 \Vec{z}=\Vec{c}, \Vec{z}\geq \Vec{0},$
	where 
\[
	A_1=	\left(
	\begin{array}{cc}
		A({\cal W})^{\top} & -A({\cal L})^{\top} \\
		 \Vec{1}^{\top} 		& -\Vec{1}^{\top} \\
	   \Vec{0}^{\top} 		& -\Vec{1}^{\top}
	\end{array}
	\right), \;\;
	\Vec{z}=
	\left( 
	\begin{array}{c} \Vec{x} \\ \Vec{y} \end{array}
	\right), 
	\mbox{ and }
	\Vec{c}=
	\left(
	\begin{array}{c}
		\Vec{0} \\ 0 \\ -1
	\end{array}
	\right). 
\]

Subsequently, we assume that D1
	is feasible.
Let $\widetilde{A_1} \Vec{z}=\widetilde{\Vec{c}}$ 
	be a linear equality system 
	obtained from  	$A_1 \Vec{z}=\Vec{c}$
	by repeatedly removing redundant equalities.
A column submatrix $\widehat{B}$ of $\widetilde{A_1}$
 	is called a {\em basis matrix}
 	if $\widehat{B}$ is a square invertible matrix.
Variables corresponding to the columns of $\widehat{B}$
	are called {\em basic variables}, 
	and $J_{\widehat{B}}$ denotes an index set of basic variables.
A  basic solution with respect to $\widehat{B}$
	is a vector $\Vec{z}$ defined by
\[
	z_i=
	\left\{ \begin{array}{ll}
		\widehat{z}_i 		& (i \in J_{\widehat{B}}), \\
		0						& (i \not \in J_{\widehat{B}}),
	\end{array} \right. 
\]
	where $\widehat{\Vec{z}}$ is a vector 
	of basic variables satisfying 
	$\widehat{\Vec{z}}=\widehat{B}^{-1}\widetilde{\Vec{c}}$.
It is well-known that 
	if a linear inequality system D1
	is feasible, 
	then it has a basic feasible solution.


Let $\Vec{z}'$ be a basic feasible solution of D1
	with respect to a basis matrix~$B$.
By Cramer's rule, 
	$z'_i =\deT(B_i)/\deT(B)$ for each $i \in J_B,$ 
	where $B_i$ is a matrix formed by replacing 
	$i$-th column of $B$ by $\widetilde{\Vec{c}}$.
Because $B_i$ is an integer matrix,
	$\deT(B)z'_i = \deT(B_i)$ is an integer for any $i \in J_B$.
Let 	
	$(\Vec{x}'^{\top}, \Vec{y}'^{\top})^{\top}$ 
	be a vector corresponding to $\Vec{z}',$
	and  $(\Vec{x^*}^{\top}, \Vec{y^*}^{\top})
		=|\deT (B)| (\Vec{x}'^{\top}, \Vec{y}'^{\top})$.
Cramer's rule states that both $\Vec{x}^*$ and $\Vec{y}^*$ are
	integer vectors.
The pair of vectors $\Vec{x}^*$ and $\Vec{y}^*$ 
	satisfies the following conditions:
\begin{eqnarray*}
	A({\cal W})^{\top}\Vec{x^*} - A({\cal L})^{\top} \Vec{y^*}
		&=&|\deT (B)| (A({\cal W})^{\top}\Vec{x}' - A({\cal L})^{\top} \Vec{y}')
		=|\deT (B)|\Vec{0}=\Vec{0},  \\
	\sum_{S \in {\cal W}} x^*_S- \sum_{S \in {\cal L}}y^*_S
		&=&|\deT (B)|(\Vec{1}^{\top}\Vec{x}'-\Vec{1}^{\top}\Vec{y}')
		=|\deT (B)| 0=0, \\
	\sum_{S \in {\cal L}}y^*_S 
		&=&|\deT (B)| \Vec{1}^{\top}\Vec{y}' =|\deT (B)|,  \\
	\Vec{x^*}& =& |\deT (B)|\Vec{x}' \geq \Vec{0},
		\; \mbox{ and  } \;
	\Vec{y^*} = |\deT (B)|\Vec{y}' \geq \Vec{0}.	
\end{eqnarray*}
Next, we construct a trading transform 
	corresponding to the pair of $\Vec{x}^*$ and $\Vec{y}^*$.
Let ${\cal X}=(X_1,X_2,\ldots ,X_{|\deT (B)|})$ be a 
	sequence of winning coalitions, where
	each winning coalition $S \in {\cal W}$
	appears in ${\cal X}$  $x^*_S$-times.
Similarly, we introduce a sequence 
	${\cal Y}=(Y_1,Y_2,\ldots ,Y_{|\deT (B)|}),$
	where each losing coalition $S \in {\cal L}$
	appears in ${\cal Y}$  $y^*_S$-times.
The above equalities imply that 
	$({\cal X};{\cal Y})$ is a trading transform of size $|\deT (B)|$.
Therefore,
	we have shown that if D1 is feasible,
	then a given simple game $G=(N, {\cal W})$ is not 
	$|\deT (B)|$-trade robust.

Finally, we provide an upper bound on $|\deT (B)|$.
Let $\alpha_n$ be the maximum of the determinants
	of  $n \times n$ 0--1 matrices.
For any  $n \times n$ 0--1 matrix $M,$
	it is easy to show that $\deT (M) \geq -\alpha_n$
	by swapping two rows of $M$ (when $n \geq 2$).    
If a column of $B$ is indexed by a component of $\Vec{x}$
	(i.e., indexed by a winning coalition), 
	then each component of the column is either $0$ or $1$.
Otherwise, a column (of $B$) 
	is indexed by a component of $\Vec{y}$
	(i.e., indexed by a losing coalition) 
	whose components are either $0$ or $-1$.
Now, we apply elementary matrix operations to $B$
	(see Figure~\ref{Fig:EMO1}).
For each column of $B$ indexed by a component of $\Vec{y}$,
	we multiply the column by $(-1)$.
The resulting matrix, denoted by $B'$, 
	is a 0--1 matrix satisfying $|\deT (B)|=|\deT (B')|$.

\begin{figure}[htb]
\[
\begin{array}{|rrrr|rr|}
\hline
\;\;0 &\;\; 0 &\;\; 1 &\;\;1 & 0 & -1 \\
0 & 1 & 0 & 1 & 0  & 0 \\
1 & 0 & 0 & 1 & 0 & -1 \\
1 & 1 & 1 & 0 &-1 & -1 \\
\hline
1 & 1 & 1 & 1 &-1 & -1 \\
\hline
0 & 0 & 0 & 0 & -1 & -1\\
\hline
\multicolumn{6}{c}{B}
\end{array}  
\quad
\begin{array}{|rrrr|rr|}
\hline
\;\;0 &\;\; 0 &\;\; 1 &\;\;1 &\;\; 0 &\;\; 1 \\
0 & 1 & 0 & 1 & 0  & 0 \\
1 & 0 & 0 & 1 & 0 & 1 \\
1 & 1 & 1 & 0 & 1 & 1 \\
\hline
1 & 1 & 1 & 1 & 1 & 1 \\
\hline
0 & 0 & 0 & 0 & 1 & 1\\
\hline
\multicolumn{6}{c}{B'}
\end{array}  
\]
\caption{Example of elementary matrix operations for D1.} \label{Fig:EMO1}
\end{figure}

As $B$ is a submatrix of $A_1$, 
	the number of rows (columns) 
	of $B$, denoted by $n'$, 
	is less than or equal to $n+2$.
When $n' < n+2$, 
	we obtain the desired result: 
	$|\deT (B)|=|\deT (B')| \leq \alpha_{n'} \leq \alpha_{n+1}$.
If  $n'=n+2$,  
	then  $B$ has a row vector corresponding to equality
	$\Vec{1}^{\top}\Vec{x}-\Vec{1}^{\top}\Vec{y}=0$, 
	which satisfies the condition that each component 
	is either $1$ or $-1$,
	and thus $B'$ has an all-one row vector.
Lemma~\ref{all-one}~(c1) appearing below	
	states that 
	$|\deT (B)|=|\deT (B')| \leq \alpha_{n'-1}\leq \alpha_{n+1}$.

\begin{lemma}\label{all-one}
Let $M$ be an $n \times n$ 0--1 matrix, 
	where $n \geq 2$.
\begin{description}
\item[(c1)] If  a row (column) vector of $M$ is the all-one vector,
	then \mbox{$|\deT (M)| \leq \alpha_{n-1}$.} 
\item[(c2)] If a row (column) vector of $M$ is a 0--1 vector 
	consisting of  
	a unique  0-component and $n-1$ 1-components, 
	then  $|\deT (M)| \leq 2\alpha_{n-1}$. 
\end{description}
\end{lemma}
	  
%
\noindent 
Proof of  (c1).  
Assume that the first column of $M$ is the all-one vector.
We apply the following elementary matrix operations to $M$
	(see Figure~\ref{Fig:EMOc1}).
For each column  of $M$
	except the first column,
	if the first component is equal to $1$, 
	then we multiply the column by $(-1)$ and 
	add the all-one column vector.
The obtained matrix, denoted by $M'$, 
	is an $n \times n$ 0--1 matrix satisfying 
	$|\deT (M)|=|\deT (M')|,$
	and the first row is a unit vector.
Thus, it is obvious that
	$|\deT (M')| \leq \alpha_{n-1}$.
	
\begin{figure}[htb]
\[
\begin{array}{|r|rrrr|}
\hline
\;\;1 &\;\; 1 &\;\; 0 &\;\; 1 &\;\; 0 \\
1 & 1 & 1 & 1 & 0 \\
1 & 0 &  1& 0 & 0 \\
1 & 1 & 1 & 0 & 1\\
1 & 0 & 0 & 1 & 1 \\
\hline
\multicolumn{5}{c}{M}
\end{array}  
\quad
\begin{array}{|r|rrrr|}
\hline
\;\;1 &\;\; 0 &\;\; 0 &\;\;0 &\;\; 0 \\
\hline
1 & 0 &1 &0 & 0 \\
1 & 1 &1 & 1 &0 \\
1 & 0 & 1&1 &1 \\
1 & 1 & 0& 0 & 1\\
\hline
\multicolumn{5}{c}{M'}
\end{array}  
\]
\caption{Example of elementary matrix operations for (c1).} \label{Fig:EMOc1}
\end{figure}
\smallskip

%

\noindent
Proof of (c2).  Assume that the first column vector of $M$, denoted by $\Vec{a}$,  
	contains exactly one 0-component.
	Obviously,  $\Vec{e}=\Vec{1}-\Vec{a}$ is a unit vector.
	Let $M_1$ and $M_e$ be a pair of matrices obtained from $M$ 
	with the first column replaced by $\Vec{1}$ and $\Vec{e}$, respectively.
Then, it is easy to prove that
\[
	|\deT (M)|=|\deT (M_1)-\deT (M_e)|\leq 
		|\deT (M_1)|+|\deT (M_e)| \leq 2 \alpha_{n-1}.
\]
\QED

\smallskip

From the above discussion,
	we obtain the following theorem
	(without the assumption 
	of the monotonicity property~(\ref{monotonicity})).
\begin{theorem} \label{non-weighted-tt}
A given simple game $G=(N, {\cal W})$ with $n$ players
	is weighted if and only if 
	$G$ is $\alpha_{n+1}$-trade robust,
	where $\alpha_{n+1}$ is the maximum of determinants
	of  $(n+1) \times (n+1)$ 0--1 matrices.
\end{theorem}

\noindent 
Proof. 
If a given simple game is not $\alpha_{n+1}$-trade robust,
	then it is not trade robust and, thus, not weighted, 
	as shown by~\cite{taylor1992characterization,taylor1999simple}. 
We have discussed the inverse implication: 
	if a given simple game $G$ is not weighted, 
	then the linear inequality system P1
	is infeasible.
Farkas' lemma~\citep{farkas1902theorie}
	implies that D1 is feasible.
From the above discussion, we have a trading transform 
	$(X_1,\ldots ,X_j;Y_1, \ldots Y_j)$
	satisfying 
	$j \leq \alpha_{n+1}$,
	$X_1, \ldots ,X_j \in {\cal W}$,
	and  $Y_1, \ldots , Y_j \in {\cal L}$.
\QED

\medskip

Applying the Hadamard's 
	evaluation~\citep{hadamard1893resolution} 
	of the determinant, 
	we obtain Theorem~\ref{HadamardUB}.
	
\begin{theorem}\label{HadamardUB}
For any positive integer $n$, 
	$\alpha_n \leq 	(n+1)^{\frac{n+1}{2}}(1/2)^{n}$.
\end{theorem}

\noindent
The exact values of $\alpha_n$ for small positive integers
	$n$ appear 
	in ``The On-Line Encyclopedia 
	of Integer Sequences (A003432)''~\citep{OEISA003432}
	and Table~\ref{Table:Exact}.

\section{Integer Weights 
	of Weighted Simple Games}\label{integerweights}

This section reviews the integer-weight 
	representations of weighted simple games.
Throughout this section, we do not need to assume 
	the monotonicity property~(\ref{monotonicity}),	
	except in Table~\ref{Table:Exact}.

\begin{theorem} \label{theorem-weightedness}
Assume that a given simple game $G=(N, {\cal W})$ satisfies 
	$\emptyset \not \in {\cal W} \ni N$.
If a given simple game $G$ is weighted, 
	then there exists an integer-weight representation
	$(q; \Vec{w}^{\top})$ of $G$  satisfying  
	$|w_i| \leq \alpha_n$ $(\forall i \in N)$, 
	$|q| \leq \alpha_{n+1}$, and 
	$1 \leq \sum_{i \in N} w_i  \leq 2 \alpha_{n+1}-1$.
\end{theorem}

\noindent
Proof.
It is easy to show that
	a given simple game $G=(N, {\cal W})$ is weighted
	if and only if the following linear inequality system is feasible:
\begin{eqnarray*}
\mbox{P2: }\;\;
	 A({\cal W}) \Vec{w} & \geq & q	 \Vec{1},\\
	 A({\cal L}) \Vec{w} & \leq & q \Vec{1}- \Vec{1},\\
	 \Vec{1}^{\top} \Vec{w} & \leq  & u-1. 
\end{eqnarray*}
We define 
\[
	A_2=\left( \begin{array}{rrr}
		 A({\cal W}) &  \Vec{1} & 0\\
		-A({\cal L}) & -\Vec{1} & 0\\
		- \Vec{1}^{\top} & 0    & 1
	\end{array}\right), \;\;
	\Vec{v}=\left( \begin{array}{r}
		\Vec{w} \\ -q \\ u
	\end{array} \right), \;\;
	\Vec{d}=\left( \begin{array}{r}
		\Vec{0} \\ \Vec{1} \\ 1
	\end{array} \right),
\]
	and denote the inequality system P2 
	by $A_2 \Vec{v} \geq \Vec{d}$.

Subsequently, we assume that P2 is feasible.
A non-singular submatrix $\widehat{B}$ of $A_2$ is called 
	a {\em basis matrix}. 
Variables corresponding to columns of $\widehat{B}$
	are called {\em basic variables}, 
	and $J_{\widehat{B}}$ denotes an index set of basic variables.
Let $\Vec{d}_{\widehat{B}}$ be a subvector 
	of $\Vec{d}$ corresponding to rows of $\widehat{B}$.
A  {\em basic solution} with respect to $\widehat{B}$
	is a vector $\Vec{v}$ defined by
\[
	v_i=
	\left\{ \begin{array}{ll}
		\widehat{v}_i 	& (i \in J_{\widehat{B}}), \\
		0						& (i \not \in J_{\widehat{B}}),
	\end{array} \right. 
\]
	where $\widehat{\Vec{v}}$ is a vector 
	of basic variables satisfying 
	$\widehat{\Vec{v}}
	=\widehat{B}^{-1}\Vec{d}_{\widehat{B}}$.
It is well-known that 
	if a linear inequality system P2 is feasible, 
	there exists a basic feasible solution.

Let $(\Vec{w}'^{\top}, -q', u')^{\top}$ 
	be a basic feasible solution of P2 
	with respect to a basis matrix $B$.
Assumption $\emptyset \not \in {\cal W}$ implies that
	$0 \leq q' -1$ and, thus, $-q' \neq 0$.
As $N \in {\cal W}$, 
	we have inequalities
	$u'-1 \geq \Vec{1}^{\top} \Vec{w}' \geq q' \geq 1,$
	which imply that $u'\neq 0$.
The definition of a basic solution  implies that
	$-q$ and $u$ are basic variables with respect to 
	the basis matrix $B$.
Thus,  $B$ has columns corresponding to 
	basic variables $-q$ and $u$.
A column of $B$ indexed by $u$ is called the last column.
As $B$ is invertible, the last column of $B$ is not the zero vector, 
	and thus $B$ includes a row 
	corresponding to inequality  $\Vec{1}^{\top}\Vec{w} \leq u-1$, 
	which is called the last row 
	 (see Figure~\ref{Fig:EMO2}).
Here,  the number of rows (columns) of $B$, denoted by $n'$, 
	is less than or equal to $n+2$.

For simplicity, we denote the basic feasible solution  
	$(\Vec{w}'^{\top}, -q', u')^{\top}$ by  $\Vec{v}'$.
By Cramer's rule, 
	$v'_i =\deT(B_i)/\deT(B)$ for each $i \in J_B,$ 
	where $B_i$ is obtained from $B$ 
	with a column corresponding to variable $v_i$ 
	replaced by $\Vec{d}_{B}$.
Because $B_i$ is an integer matrix,
	$\deT(B)v'_i = \deT(B_i)$ is an integer for any $i \in J_B$. 
Cramer's rule states that $(\Vec{w^*}^{\top}, -q^*, u^*)
	=|\deT (B)| (\Vec{w}'^{\top}, -q', u')$ 
	is an integer vector satisfying the following conditions:
\begin {eqnarray*}
A({\cal W})\Vec{w^*} &=& |\deT (B)| A({\cal W})\Vec{w}'
		\geq |\deT (B)| q'\Vec{1} =q^*\Vec{1}, \\
A({\cal L})\Vec{w^*} &=&  |\deT (B)| A({\cal L})\Vec{w}'
		\leq  |\deT (B)| (q'\Vec{1}-\Vec{1})
		\leq q^*\Vec{1} -\Vec{1}, \;\; \mbox{and} \\
\Vec{1}^{\top}\Vec{w^*} &=&  |\deT (B)|\Vec{1}^{\top}\Vec{w}'
		\leq |\deT (B)| (u'-1) \leq u^* -1.
\end{eqnarray*}
From the above, $(q^*; \Vec{w^*}^{\top})$ 
	is an integer-weight representation of $G$.
As $N \in {\cal W}$, 
	we obtain 
	$\Vec{1}^{\top}\Vec{w^*} \geq q^*=|\deT (B)|q'\geq 1$.
	
\begin{figure}[htb]
{\scriptsize 
\[
\begin{array}{l@{\quad}l@{\quad}l@{\quad}l}
\begin{array}{|rrrr|r|r|}
\multicolumn{1}{r}{w_1} & w_2 & w_3 & 
\multicolumn{1}{r}{w_4} &
\multicolumn{1}{r}{-q} &
\multicolumn{1}{r}{\;\;u} \\
\hline
1 & 1 & 1 & 0 & 1 & 0 \\
0 & 1 & 1 & 1 & 1 & 0 \\
\hline
0 &-1&-1 & 0 &-1 & 0 \\
-1& 0 & 0 &-1 &-1 & 0 \\
0 &-1 & 0 &-1 &-1 & 0 \\
\hline
-1&-1 &-1 &-1 & 0 & 1 \\
\hline
\multicolumn{6}{c}{B}
\end{array}
\\ \\
\begin{array}{|rrrr|r|r|}
\multicolumn{1}{r}{w_1} & w_2 & w_3 & 
\multicolumn{1}{r}{w_4} &
\multicolumn{1}{r}{-q} &
\multicolumn{1}{r}{\;\;u} \\
\hline
1 & 1 & 1 & 0 & 0 & 0 \\
0 & 1 & 1 & 1 & 0 & 0 \\
\hline
0 &-1&-1 & 0 & 1 & 0 \\
-1& 0 & 0 &-1 &1 & 0 \\
0 &-1 & 0 &-1 & 1 & 0 \\
\hline
-1&-1 &-1 &-1 & 1 & 1 \\
\hline
\multicolumn{6}{c}{B_q}
\end{array}
&
\begin{array}{|rrrr|r|}
\multicolumn{1}{r}{w_1} & w_2 & w_3 & 
\multicolumn{1}{r}{w_4} &
\multicolumn{1}{r}{-q}  \\
\hline
1 & 1 & 1 & 0 & 0  \\
0 & 1 & 1 & 1 & 0  \\
\hline
0 &-1&-1 & 0 & 1  \\
-1& 0 & 0 &-1 &1  \\
0 &-1 & 0 &-1 & 1  \\
\hline 
\multicolumn{5}{c}{} \\
\multicolumn{5}{c}{B'_q}
\end{array}
&
\begin{array}{|rrrr|r|}
\multicolumn{1}{r}{w_1} & w_2 & w_3 & 
\multicolumn{1}{r}{w_4} &
\multicolumn{1}{r}{-q}  \\
\hline
1 & 1 & 1 & 0 & 0  \\
0 & 1 & 1 & 1 & 0  \\
\hline
0 &1 & 1 & 0 & 1  \\
1 & 0 & 0 & 1 &1  \\
0 &1 & 0 & 1 & 1  \\
\hline 
\multicolumn{5}{c}{} \\
\multicolumn{5}{c}{B''_q}
\end{array}
\\ \\
\begin{array}{|r|r|rr|r|r|}
\cline{2-2}
\multicolumn{1}{r|}{w_1} & w_2 & w_3 & 
\multicolumn{1}{r}{w_4} &
\multicolumn{1}{r}{-q} &
\multicolumn{1}{r}{\;\;u} \\
\hline
1 & 0 & 1 & 0 & 1 & 0 \\
0 & 0 & 1 & 1 & 1 & 0 \\
\hline
0 & 1&-1 & 0 & -1 & 0 \\
-1& 1 & 0 &-1 &-1 & 0 \\
0 & 1 & 0 &-1 & -1 & 0 \\
\hline
-1& 1 &-1 &-1 & 0 & 1 \\
\hline
\multicolumn{6}{c}{B_2}
\end{array}
&
\begin{array}{|r|r|rr|r|}
\cline{2-2}
\multicolumn{1}{r|}{w_1} & w_2 & w_3 & 
\multicolumn{1}{r}{w_4} &
\multicolumn{1}{r}{-q} \\
\hline
1 & 0 & 1 & 0 & 1  \\
0 & 0 & 1 & 1 & 1  \\
\hline
0 & 1&-1 & 0 & -1  \\
-1& 1 & 0 &-1 &-1  \\
0 & 1 & 0 &-1 & -1  \\
\hline
\multicolumn{5}{c}{} \\
\multicolumn{5}{c}{B'_2}
\end{array}
&
\begin{array}{|r|r|rr|r|}
\cline{2-2}
\multicolumn{1}{r|}{w_1} & w_2 & w_3 & 
\multicolumn{1}{r}{w_4} &
\multicolumn{1}{r}{-q} \\
\hline
1 & 0 & 1 & 0 & 1  \\
0 & 0 & 1 & 1 & 1  \\
\hline
0 & 1& 1 & 0 & 1  \\
1& 1 & 0 & 1 &1  \\
0 & 1 & 0 & 1 & 1  \\
\hline
\multicolumn{5}{c}{} \\
\multicolumn{5}{c}{B''_2}
\end{array}
\\ \\
\begin{array}{|rrrr|r|r|}
\multicolumn{1}{r}{w_1} & w_2 & w_3 & 
\multicolumn{1}{r}{w_4} &
\multicolumn{1}{r}{-q} &
\multicolumn{1}{r}{\;\;u} \\
\hline
1 & 1 & 1 & 0 & 1 & 0 \\
0 & 1 & 1 & 1 & 1 & 0 \\
\hline
0 &-1&-1 & 0 &-1 & 1 \\
-1& 0 & 0 &-1 &-1 & 1 \\
0 &-1 & 0 &-1 &-1 & 1 \\
\hline
-1&-1 &-1 &-1 & 0 & 1 \\
\hline
\multicolumn{6}{c}{B_u}
\end{array}
&
\begin{array}{|rrrr|r|r|}
\multicolumn{1}{r}{w_1} & w_2 & w_3 & 
\multicolumn{1}{r}{w_4} &
\multicolumn{1}{r}{-q} &
\multicolumn{1}{r}{\;\;u} \\
\hline
1 & 1 & 1 & 0 & 1 & 0 \\
0 & 1 & 1 & 1 & 1 & 0 \\
\hline
0 &1 & 1 & 0 &1 & 1 \\
1 &0 & 0 & 1 &1 & 1 \\
0 &1 & 0 & 1 &1 & 1 \\
\hline
1 &1 &1  &1  & 0 & 1 \\
\hline
\multicolumn{6}{c}{B'_u}
\end{array}
\end{array}
\]
}
\caption{Examples of elementary matrix operations for P2.} \label{Fig:EMO2}
\end{figure}

Now, we discuss the magnitude of $|q^*|=|\deT (B_q)|,$
	where $B_q$ is obtained from $B$ with a column 
	corresponding to variable $-q$ replaced by $\Vec{d}_B$.
As the last column of $B_q$ is a unit vector,  
	we delete the last column and the last row from $B_q$ and
	obtain a matrix $B'_q$ satisfying $\deT (B_q)=\deT (B'_q)$.
We apply the following elementary matrix operations to  $B'_q$.
First, we multiply the column corresponding to variable $-q$ 
	(which is equal to $\Vec{d}_B$) by $(-1)$.
Next, we multiply the rows indexed by losing coalitions by $(-1)$.
The resulting matrix, denoted by $B''_q$, 
	is 0--1 valued and satisfies the following condition:
\[
	|q^*| =|\deT (B_q)|=|\deT (B'_q)|=|\deT (B''_q)|
	\leq \alpha_{n'-1} \leq \alpha_{n+1}.
\]

Next, we show that  $|w^*_i| \leq \alpha_n$ $(i \in N)$.
If $w^*_i \neq 0$, then $w_i$ is a basic variable 
	that satisfies $|w^*_i|=| \deT (B_i)|,$
	where $B_i$ is obtained from $B$ but the column
	corresponding to variable $w_i$ is replaced by $\Vec{d}_B$.
In a manner similar to that above, 
	we  delete the last column and the last row from $B_i$ and
	obtain a matrix $B'_i$ satisfying $\deT (B_i)=\deT (B'_i)$.
Next, we multiply a column corresponding to variable $w_i$ 
	by $(-1)$.
We multiply rows  indexed by losing coalitions
	by $(-1)$
	and obtain a 0--1 matrix $B''_i$.
Matrix $B_i$ contains a column corresponding to the original 
	variable $-q$, which contains values  $1$ or $-1$.
Thus,  matrix $B''_i$ contains a column vector 
	that is equal to an all-one vector.
Lemma~\ref{all-one}~(c1) implies that 
\[
		|w^*_i|=|\deT (B_i)|=|\deT (B'_i)|
	=|\deT (B''_i)| \leq \alpha_{n'-2} \leq \alpha_n. 
\]

Lastly, we discuss the value of $|u^*|=|\deT (B_u)|,$
	where $B_u$ is obtained from $B$ but the last column
	(column indexed by variable $u$) is replaced by $\Vec{d}_B$.
In a manner similar to that above, 
	we multiply the last column by $(-1)$,  
	multiply the rows  indexed by losing coalitions by $(-1)$, 
	and multiply the last row by $(-1)$.
The resulting matrix,  denoted by $B'_u$, 
	is a 0--1 matrix in which the last row contains 
	exactly one 0-component (indexed by variable $-q$).
Lemma~\ref{all-one}~(c2) implies that
\[
	  |u^* |=|\deT (B_u)|=|\deT (B'_u)| \leq 2 \alpha_{n'-1} \leq 2 \alpha_{n+1},  
\]
and thus 
  $\Vec{1}^{\top}\Vec{w^*}
  	\leq  u^*-1 \leq |u^*| -1 
  	\leq 2 \alpha_{n+1}-1$.\QED 

\medskip

%
\cite{kurz2012minimum} exhaustively generated 
	all weighted voting games 
	satisfying the monotonicity property~(\ref{monotonicity})
 	for up to nine voters.
Table~\ref{Table:Exact} shows maxima of the exact values 
	of minimal integer-weight representations obtained 
	by~\citep{kurz2012minimum}, 
	Muroga's bounds in~\cite{muroga1971threshold}, 
	and our upper bounds.
The table shows that our bound on the quota is tight
	when $n \leq 5$.

\begin{table}[htb]
\caption{Exact values of integer weights representations.}
\label{Table:Exact}
\begin{tabular}{@{}c@{}|rrrrrrrrrrr@{}}
\hline
 $n$&1&2&3&4&5&6&7&8&9&10&11 \\ 
\hline
 $\alpha_n \;\; \dagger $ 
  &1&1&2&3&5&9&32&56&144&320&1458 \\
\hline
$\displaystyle 
	\max_{\scriptsize (N, {\cal W})}
	\min_{\scriptsize [q; w]} 
	\max_i w_i \;\; \ddagger $  
&1&1&2&3&5&9&18&42&110 \\[-0.8ex]
{\small Muroga's bound}  $( \alpha_n ) \bullet$  &1&1&2&3&5&9&32&56&144&320&1458 \\
\hline
$\displaystyle  
	\max_{\scriptsize (N, {\cal W})}
	\min_{\scriptsize [q; w]} \;q \;\;   
	\ddagger  $  
&1&2&3&5&9&18&40&105&295 \\[-0.8ex]
{\small Our bound} $( \alpha_{n+1} )$ &1&2&3&5&9&32&56&144&320&1458 \\
{\small Muroga's bound} $(n\alpha_n) \bullet$ &
	1&	2&	6&	12&25&	54&	224&	448&	1296&	3200&	16038 \\
\hline
$\displaystyle 
	\max_{\scriptsize (N, {\cal W})}
	\min_{\scriptsize [q; w]} 
	{\textstyle \sum_i w_i} \;\; \ddagger $ &
	1&2&4&8&15&33&77&202&568 \\
{\small Our bound} $(2\alpha_{n+1}-1)$ 
& 	1&3&5&9&17&63&111&287&639&2915  \\
\hline
\end{tabular}

$\dagger$ \cite{OEISA003432},  
$\ddagger$ \cite{kurz2012minimum},
$\bullet$ \cite{muroga1971threshold}.
\end{table}

\section{Rounding Method}\label{approximation}

This section addresses the problem
	of finding integer-weight representations.
In this section, 
	we assume the monotonicity property~(\ref{monotonicity}).
In addition, a weighted simple game is
	given by a triplet $(N,\minW, \maxL),$
	where $\minW$ and  $\maxL$ denote
	 the set of minimal winning coalitions
	and the set of maximal losing coalitions, respectively.
We also assume that the empty set is a losing coalition,
	$N$ is a winning coalition, 
	and every player in $N$ is not 
	a null player.
Thus, there exists 
	an integer-weight representation
	in which $q \geq 1$ and 
	$w_i \geq 1 \; (\forall i \in N)$.

We discuss a problem 
	for finding an integer-weight representation,
	which is formulated 
	by the following integer programming problem:
\begin{eqnarray}
\nonumber \mbox{Q: }  \mbox{find a vector } &&(q; \Vec{w}) \\
\mbox{satisfying} && 
		\sum_{i \in S} w_i \geq q \;\;\; (\forall S \in \minW), 
			\label{Winning} \\
&&  \sum_{i \in S} w_i \leq q-1 \;\;\; (\forall S \in \maxL), 
			\label{Losing} \\
&&  q \geq 1, \;\;\; w_i \geq 1 \;\;\; (\forall i \in N), 
			\label{one} \\
&&  q \in \Int, \;\;\;  w_i \in \Int \;\;\; (\forall i \in N).
			\label{integer} 
\end{eqnarray}
%
%
%
A linear relaxation problem \overQ \, is obtained 
	from Q by dropping the integer constraints~(\ref{integer}). 

Let $(q^*; {\Vec{w}^*}^{\top})$ 
	be a basic feasible solution of the linear inequality system~\overQ.
Our proof in the previous section showed that 
$ |\deT (B^*)|(q^*;  {\Vec{w}^*}^{\top})$
	gives a solution of Q 
	(i.e., an integer-weight representation), 
	where $B^*$ denotes a corresponding basis matrix 
	of  \overQ.
%
When $|\deT (B^*)| >n$, 
	there exists a simple method for generating 
	a smaller integer-weight representation. 
For any weight vector $\Vec{w}=(w_1, w_2, \ldots, w_n)^{\top}$,
	we denote the integer vector
	$(\lfloor w_1 \rfloor, \lfloor w_2 \rfloor, \ldots ,
		\lfloor w_n \rfloor)^{\top}$ by 
	$\lfloor  \Vec{w} \rfloor$.
Given a solution $(q^*; {\Vec{w}^*}^{\top})$ of~\overQ,
	we introduce an integer vector  
	$\Vec{w}'=\lfloor n \Vec{w}^* \rfloor$
	and an integer $q'=\lfloor n(q^*-1)\rfloor +1 $.
For any minimal winning coalition $S \in \minW$,
	we have that
\begin{eqnarray*}
	\sum_{i \in S}w'_i &>&  \sum_{i \in S} (n w^*_i -1)
	\geq n \sum_{i \in S}w^*_i -n 
	\geq nq^* -n = n(q^*-1) \geq \lfloor n(q^*-1) \rfloor,  \\
	\sum_{i \in S}w'_i 
		&\geq & \lfloor n(q^*-1) \rfloor +1 = q'.
\end{eqnarray*}
Each maximal losing coalition $S \in \maxL$ satisfies
\begin{eqnarray*}
	\sum_{i \in S}w'_i &\leq & \sum_{i \in S} nw^*_i
	\leq n(q^*-1), \\
 \sum_{i \in S}w'_i &\leq & \lfloor n(q^*-1) \rfloor 
	=q'-1.  
\end{eqnarray*}
Thus, the pair of $\Vec{w}'$ and $q'$ gives 
	an integer-weight representation 
	satisfying $(q'; \Vec{w}'^{\top}) \leq n (q^*; {\Vec{w}^*}^{\top})$.
In the remainder of this section,
	we show that there exists an integer-weight representation (vector)
	that  is less than or equal to
	$	 ((2-\sqrt{2})n+(\sqrt{2}-1))
		 (q^*; {\Vec{w}^*}^{\top})
		 < (0.5858n+0.4143) (q^*; {\Vec{w}^*}^{\top})$
		for any solution $(q^*; {\Vec{w}^*}^{\top})$ of \overQ.
%

\begin{theorem}\label{bullet}
Let $(q^*; {\Vec{w}^*}^{\top})$ be a solution of \overQ.
We define $\ell_1=(2-\sqrt{2})n-(\sqrt{2}-1)$ and
	$u_1=(2-\sqrt{2})n+(\sqrt{2}-1)$.
Then, there exists a real number 
	$\lambda^{\bullet}  \in [\ell_1, u_1]$ 
	so that the pair
	 $Q=\lfloor \lambda^{\bullet} (q^*-1) \rfloor +1$
	and $\Vec{W}=\lfloor \lambda^{\bullet} \Vec{w}^* \rfloor $ 
	gives a feasible solution of Q (i.e., an integer-weight representation).
	 
\end{theorem}

\noindent
Proof.
For any positive real $\lambda$, it is easy to see that
	each maximal losing coalition $S \in \maxL$ satisfies 
\begin{eqnarray*}
	 \sum_{i \in S} \lfloor \lambda w^*_i \rfloor  
	&\leq & \sum_{i \in S} \lambda w^*_i
	\leq \lambda (q^*-1), \\
	\sum_{i \in S} \lfloor \lambda w^*_i \rfloor  
	&\leq & 
	\lfloor \lambda (q^*-1) \rfloor.
\end{eqnarray*}

To discuss 
	the weights of minimal winning coalitions, 
	we introduce a function
$
	g(\lambda)=\lambda 
		-\sum_{i \in N}
		(\lambda w^*_i - \lfloor \lambda w^*_i \rfloor)
$.
In the second  part of this proof,
	we show that
	if we choose $\Lambda \in [\ell_1, u_1]$
	uniformly at random, 
	then 	$\Exp [g(\Lambda)] \geq 0$.
This implies that $\exists \lambda^{\bullet} \in [\ell_1, u_1]$
	satisfying $g(\lambda^{\bullet})>0$,
	because $g(\lambda)$ is right-continuous, piecewise linear, 
	and not a constant function.
When  $g(\lambda^{\bullet})>0$,
	each minimal winning coalition $S \in \minW$ satisfies 
\begin{equation}
\lambda^{\bullet} >
		\sum_{i \in N}
	(\lambda^{\bullet} w^*_i - \lfloor \lambda^{\bullet} w^*_i \rfloor)
	\geq 
	\sum_{i \in S}
	(\lambda^{\bullet} w^*_i 
		- \lfloor \lambda^{\bullet} w^*_i \rfloor)
	=	\sum_{i \in S} \lambda^{\bullet} w^*_i 
		- \sum_{i \in S} \lfloor \lambda^{\bullet} w^*_i \rfloor,  
\end{equation}
which implies
\[
	\sum_{i \in S} 
		\lfloor \lambda^{\bullet} w^*_i \rfloor
	>
	\sum_{i \in S}  \lambda^{\bullet} w^*_i -\lambda^{\bullet}
		= \lambda^{\bullet}\left( \sum_{i \in S}w^*_i -1 \right)
		\geq \lambda^{\bullet}(q^*-1) 
		\geq \lfloor \lambda^{\bullet}(q^*-1) \rfloor, 
\]
and thus
\[
	\sum_{i \in S} 
		\lfloor \lambda^{\bullet} w^*_i \rfloor
	\geq  \lfloor \lambda^{\bullet}(q^*-1) \rfloor +1. 
\]

\indent
Finally, we show that $\Exp [g(\Lambda )] \geq 0$
	if we choose $\Lambda \in [\ell_1, u_1]$
	uniformly at random.
It is obvious that
\begin{eqnarray*}
	\Exp [g(\Lambda )]&=&\Exp [\Lambda ]
		- \sum_{i \in N} 
		\Exp [(\Lambda w^*_i - \lfloor \Lambda w^*_i \rfloor)]
	=\frac{\ell_1 + u_1}{2} 
	-	 \sum_{i \in N}
	\frac{\displaystyle 
		\int_{\ell_1}^{u_1}
		(\lambda w^*_i - \lfloor \lambda w^*_i \rfloor)
		 \dd \lambda
	}{u_1-\ell_1} \\
	&=& (2-\sqrt{2}) n
	-	 \sum_{i \in N}
	\frac{\displaystyle 
		\int_{\ell_1}^{u_1}
		(\lambda w^*_i - \lfloor \lambda w^*_i \rfloor)
		 \dd \lambda
	}{u_1-\ell_1}.
\end{eqnarray*}

\begin{figure}[htbp]
 \begin{center}
  \includegraphics[width=100mm]{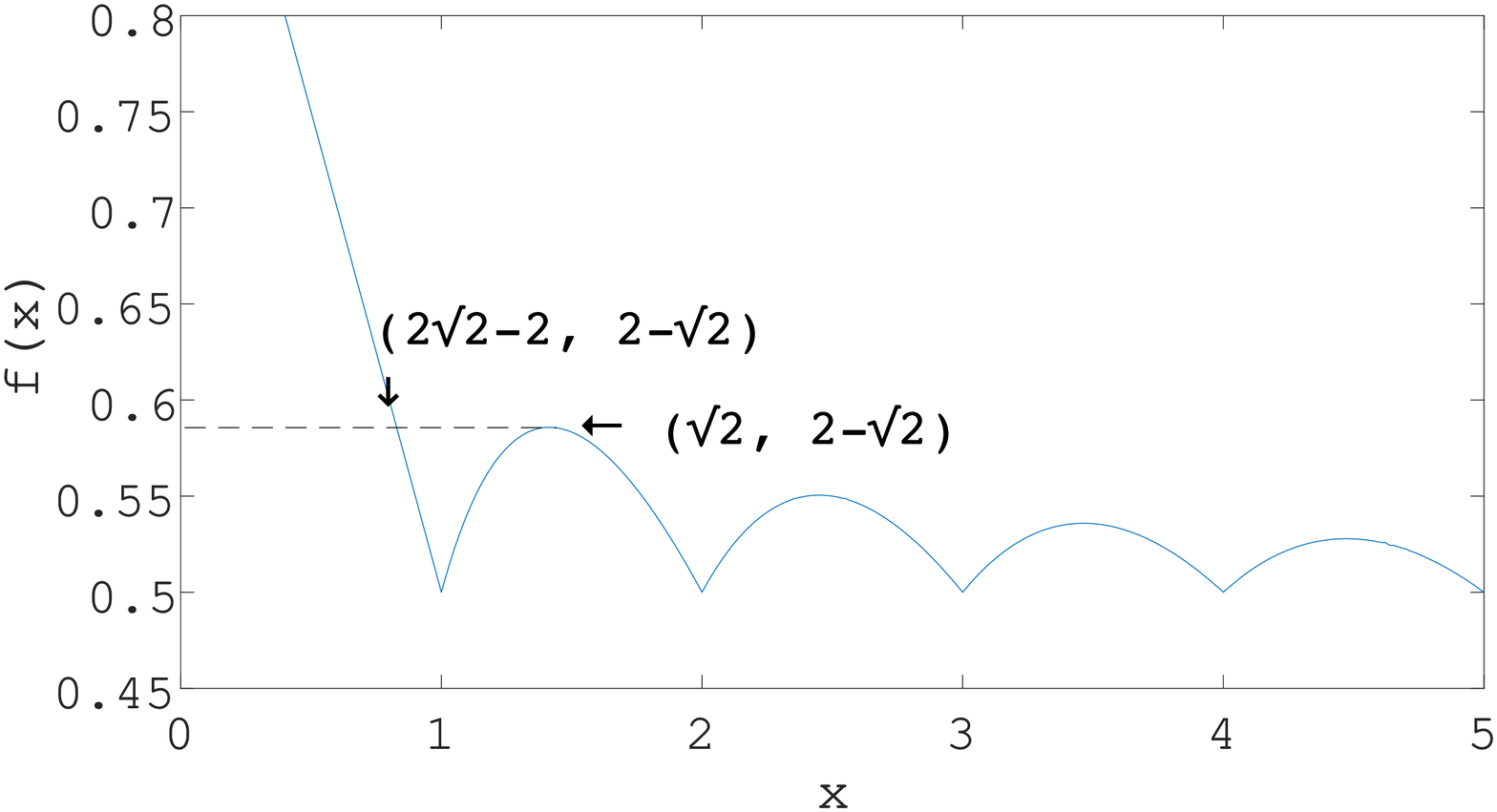}
 \end{center}
 \caption{Plot of function
 	$\displaystyle f(x) = \frac{\displaystyle
		\int_{-x}^{0}
		(\mu - \lfloor \mu \rfloor)
		 \dd \mu
	         }{x} $.  }
 \label{fig:one}
\end{figure}

\noindent
Let us discuss the last term appearing above.	 
By substituting 	 $\mu$ for $\lambda w^*_i$, we obtain 
	 
\begin{eqnarray*}
	\frac{\displaystyle
		\int_{\ell_1}^{u_1}
		(\lambda w^*_i - \lfloor \lambda w^*_i \rfloor)
		 \dd \lambda
	}{u_1-\ell_1} 
	&=&
	\frac{\displaystyle
		\int_{\ell_1 w^*_i}^{u_1 w^*_i}
		(\mu - \lfloor \mu \rfloor)
		 \dd \mu
	}{w^*_i (u_1 -\ell_1)}  \\
	&\leq &
	\frac{\displaystyle
		\int_{-w^*_i(u_1 -\ell_1)}^{0}
		(\mu - \lfloor \mu \rfloor)
		 \dd \mu
	}{w^*_i (u_1 -\ell_1)} 
	=
	\frac{\displaystyle
		\int_{-x}^{0}
		(\mu - \lfloor \mu \rfloor)
		 \dd \mu
	}{x},
\end{eqnarray*}

\noindent
where the last equality is obtained by
	setting $x=w^*_i(u_1-\ell_1)$.
As $u_1 - \ell_1 = 2(\sqrt{2} -1) $ 
	 and $w^*_i\geq 1$,
	 it is clear that $x = w^*_i(u_1-\ell_1) \geq 2(\sqrt{2}-1)$.
Here, we introduce a function  
	$\displaystyle f(x) = \frac{\displaystyle
		\int_{-x}^{0}
		(\mu - \lfloor \mu \rfloor)
		 \dd \mu
	         }{x} $.
According to numerical calculations (see Figure~\ref{fig:one}), 
	inequality $x \geq 2(\sqrt{2}-1)$ 
	implies that $f(x)\leq 2-\sqrt{2}$.
%

\noindent
From the above, we obtain the desired result
\[
	\Exp [g(\Lambda )] \geq (2-\sqrt{2})n 
		- \sum_{i \in N}(2-\sqrt{2})
	=(2-\sqrt{2})n-(2-\sqrt{2})n=0.
\] 
\QED

\medskip

\section{Roughly Weighted Simple Games} \label{RWSG}

In this section, 
	we discuss roughly weighted simple games.
First, we show an upper bound of the length of 
	a potent certificate of non-weightedness.

\begin{theorem} \label{r-nonweightedness}
	Assume that a given simple game 
	$G=(N, {\cal W})$ satisfies 
	$\emptyset \not \in {\cal W} \ni N$ and 
	the monotonicity property~(\ref{monotonicity}).	
If a given simple game $G$ is not roughly weighted,  
	then there exists a potent certificate of non-weightedness 
	whose length is less than or equal to 	
	$ 2 \alpha_{n+1}$.
\end{theorem}

\noindent
Proof. 
Let us introduce a  linear inequality system: 
\[
\begin{array}{rrr}
	\mbox{P3:}
	\left(
	\begin{array}{rr}
		 A({\cal W})&  \Vec{1}  \\
		-A({\cal L})& -\Vec{1}  
	\end{array}
	\right)
	\left(
	\begin{array}{r}
		\Vec{w} \\ - q  
	\end{array}
	\right)
	&\geq & \Vec{0}, \\
	\Vec{1}^{\top} \Vec{w} & > & 0.
\end{array} 
\]
First, we show that  
	if P3 is feasible, then a given simple game is roughly weighted.
Let  $(q'; \Vec{w}'^{\top})$ be a feasible solution of P3. 
We introduce a new voting weight $w''_i= \max \{w'_i, 0\}$
	for each  $ i \in N$.
We show that $(q'; \Vec{w}''^{\top})$ is a rough voting representation.
As $\Vec{1}^{\top} \Vec{w}'  >  0$, 
	vector $\Vec{w}'$ includes at least one positive component,
	and thus  $ \Vec{w}'' \neq \Vec{0}$. 
If a coalition $S$ satisfies $\sum_{i \in S} w''_i <q'$, 
	then $q' > \sum_{i \in S} w''_i \geq \sum_{i \in S} w'_i,$
	and thus $S$ is losing.
Consider the case in which a coalition $S$ satisfies 
   $\sum_{i \in S} w''_i > q'$.
Let $S'=\{i \in S \mid w'_i >0\}$.
It is obvious that
	$q' < \sum_{i \in S}w''_i = \sum_{i \in S'} w''_i =\sum_{i \in S'} w'_i$ 
	and thus $S'$ is winning.
The monotonicity property~(\ref{monotonicity}) 
	and $S' \subseteq S$ imply that $S$ is winning. 

From the above discussion, 
	it is obvious that 
	if a given simple game is not roughly weighted, 
	then P3 is infeasible.
Farkas' Lemma~\citep{farkas1902theorie} 
	states that  
\begin{eqnarray*}
	\mbox{D3:}
	\left(
	\begin{array}{cc}
		A({\cal W})^{\top} & -A({\cal L})^{\top} \\
		 \Vec{1}^{\top} 		& -\Vec{1}^{\top} \\
	\end{array}
	\right)
	\left(
	\begin{array}{c}
		\Vec{x}  \\ \Vec{y}
	\end{array}
	\right)
	& = & 
	\left(
	\begin{array}{c}
		- \Vec{1} \\ 0
	\end{array}
	\right), \\
	\Vec{x} \geq \Vec{0}, \;\; \Vec{y} \geq \Vec{0},  
\end{eqnarray*} 
is feasible if and only if P3 is infeasible.
By introducing an artificial non-negative variable $u\geq 0$ 
	and equality $\Vec{1}^{\top}\Vec{x}=u-1$, 
	we obtain a linear inequality system: 
\begin{eqnarray*}
	\mbox{D3$^+$:}
	\left(
	\begin{array}{ccc}
		A({\cal W})^{\top} & -A({\cal L})^{\top} & \Vec{0} \\
		 \Vec{1}^{\top} 		& -\Vec{1}^{\top} 		& 0 \\
		 \Vec{1}^{\top}		&  \Vec{0}^{\top}		& -1 \\
	\end{array}
	\right)
	\left(
	\begin{array}{c}
		\Vec{x}  \\ \Vec{y} \\ u
	\end{array}
	\right)
	& = & 
	\left(
	\begin{array}{c}
		- \Vec{1} \\ 0 \\ -1 
	\end{array}
	\right), \\
	\Vec{x} \geq \Vec{0}, \;\; \Vec{y} \geq \Vec{0}, \;\; u\geq 0.  
\end{eqnarray*}
It is obvious that D3 is feasible if and only if D3$^+$ is feasible.

Next, we construct a  trading transform 
	from a basic feasible solution of~D3$^+$.
Let $\widetilde{A_3} \Vec{z}=\widetilde{\Vec{c}}'$ 
	be a linear equality system 
	obtained from  	D3$^+$
	by repeatedly removing redundant equalities.
As D3$^+$ is feasible, 
	there exists a basic feasible solution,
	denoted by $\Vec{z}'$, 
	and a corresponding basis matrix 
	$B$ of  $\widetilde{A_3}$.
From Cramer's rule,  $z'_S \neq 0$ implies that 
	$z'_S =\deT(B_S)/\deT(B)$  for each $S \subseteq N,$ 
	where $B_S$ is obtained from $B$ with a column
	corresponding to variable $z_S$
	replaced by $\widetilde{\Vec{c}}'$.	
Obviously, $|\deT (B)| \Vec{z}'$ is a non-negative  integer vector.
We denote by $(\Vec{x}'^{\top}, \Vec{y}'^{\top}, u')^{\top} $ 
	the basic feasible solution $\Vec{z}'$. 
We recall that $\emptyset \not \in {\cal W} \ni N$
	and introduce a pair of non-negative integer vectors 
	$(\Vec{x}^*, \Vec{y}^*)$ defined as follows: 
\begin{eqnarray*}
	x^*_S&=&\left\{ \begin{array}{ll}
		|\deT (B)| x'_S 		& (\mbox{if } S \in {\cal W}\setminus \{N\}), \\
		|\deT (B)| (x'_N+1)	&(\mbox{if } S=N),
		\end{array} \right. \\
	y^*_S&=&\left\{ \begin{array}{ll}
		|\deT (B)| y'_S 		& (\mbox{if } S \in {\cal N}\setminus \{\emptyset \}), \\
		|\deT (B)| (y'_{\emptyset}+1)	&(\mbox{if } S=\emptyset).
		\end{array} \right.
\end{eqnarray*}
Subsequently,  $\chi (S)$ denotes the characteristic vector of a coalition $S$. 
It is easy to see that 
	pair  $(\Vec{x}^*, \Vec{y}^*) $ satisfies 
\begin{eqnarray*}
\lefteqn{
	A({\cal W})^{\top} \Vec{x}^* - A({\cal L})^{\top}\Vec{y}^*
	= \sum_{S\in {\cal W}} \chi (S) x^*_S 
	- \sum_{S \in {\cal L}} \chi (S) y^*_S } \\
	&=& \sum_{S \in {\cal W}}  \chi (S) |\deT (B)|x'_S 
	+\chi (N) |\deT (B)| 
	-  \sum_{S \in {\cal L}} \chi (S) |\deT (B)| y'_S  
	-   \chi (\emptyset) |\deT (B)|   \\
	&=& |\deT (B)|
		\left(
			\left(
				\sum_{S \in {\cal W}} \chi (S) x'_S
				-  \sum_{S \in {\cal L}} \chi (S)  y'_S
			\right)
			 +\chi (N)  - \chi (\emptyset)
		\right) \\
	&=& |\deT (B)| 
		\left(
			 \left( A({\cal W})^{\top}\Vec{x}'  -A({\cal L})^{\top}  \Vec{y}' \right)  
			+\Vec{1} -\Vec{0}
		\right)
	=  |\deT (B)| (-\Vec{1}+\Vec{1} -\Vec{0})=\Vec{0}
\end{eqnarray*}
and
\begin{eqnarray*}
\lefteqn{
	\sum_{S \in {\cal W}}x^*_S -\sum_{S \in {\cal L}}y^*_S
	=\sum_{S \in {\cal W}} |\deT B| x'_S + |\deT (B)|
	-\sum_{S \in {\cal L}} |\deT (B)| y'_S - |\deT (B)|} \\
	&=&  |\deT (B)| 
	\left(
		\left(
			  \sum_{S \in {\cal W}}  x'_S
			-\sum_{S \in {\cal L}}  y'_S 
		\right)
	 	+ 1- 1
	\right)
	= |\deT (B)| (0+1-1) =0.
\end{eqnarray*}
Next, we can construct a  trading transform $({\cal X}; {\cal Y})$
	corresponding to the pair of  $\Vec{x}^*$ and $\Vec{y}^*$
	by analogy with the proof of Theorem~\ref{non-weighted-tt}.
Both $x^*_N$ and $y^*_{\emptyset}$ are positive and 
	$\emptyset \not \in {\cal W} \ni N$; therefore 
	$({\cal X}; {\cal Y})$ 
	is a potent certificate of non-weightedness.

Lastly,  we discuss the length of $({\cal X}; {\cal Y})$.
The number of rows (columns) of $B$,
	denoted by $n'$, is less than or equal to $n+2$.
The basic feasible solution $\Vec{z}'$ satisfies that
	$u'=1+\Vec{1}^{\top}\Vec{x}' \geq 1 >0$, 
	and thus Cramer's rule states that $\deT (B)u'=\deT (B_u)$
	(Figure~\ref{Fig:EMO-D3}  shows an example).
We multiply columns of $B_u$ corresponding to components in 
	$(\Vec{y}^{\top}, u)$  by $(-1)$ 
	and obtain a 0--1 matrix $B'_u$
	satisfying $|\deT (B_u)|=|\deT (B'_u)|$.	
As $\widetilde{\Vec{c}}'$ includes at most one 0-component, 
	Lemma~\ref{all-one} implies that 
	$|\deT (B'_u)| \leq 2 \alpha_{n'-1} \leq 2 \alpha_{n+1}$. 
Thus, the length of $({\cal X}; {\cal Y})$ satisfies 
\begin{eqnarray*}
	 \sum_{S \in {\cal W}} x^*_S 
	&=& \sum_{S \in {\cal W}} |\deT (B)|x'_S + |\deT (B)| 
	=|\deT (B)| ( \Vec{1}^{\top} \Vec{x}' +1 )   \\
	&=& |\deT (B)|(u'-1+1) = |\deT (B)|u' 
	= |\deT (B) u'| \\
	&=&|\deT (B_u)| =|\deT (B'_u)| \leq 2 \alpha_{n+1}. 
\end{eqnarray*}
\QED

\begin{figure}[htb]
\[
\begin{array}{ll}
\begin{array}{|rrrr|rr|r|}
\multicolumn{7}{r}{u} \\
\hline
\;\;0 &\;\; 0 &\;\; 1 &\;\;1 & 0 & -1 & 0 \\
0 & 1 & 0 & 1 & 0  & 0 & 0 \\
1 & 0 & 0 & 1 & 0 & -1 & 0 \\
1 & 1 & 1 & 0 &-1 & -1 & 0 \\
\hline
1 & 1 & 1 & 1 &-1 & -1 & 0 \\
\hline
1 & 1 & 1 & 1 & 0 & 0 & -1 \\
\hline
\multicolumn{6}{c}{B}
\end{array}  
\\ \\
\begin{array}{|rrrr|rr|r|}
\multicolumn{7}{r}{\widetilde{\Vec{c}}'} \\
\hline
\;\;0 &\;\; 0 &\;\; 1 &\;\;1 & 0 & -1 & -1 \\
0 & 1 & 0 & 1 & 0  & 0 & -1 \\
1 & 0 & 0 & 1 & 0 & -1 & -1 \\
1 & 1 & 1 & 0 &-1 & -1 & -1 \\
\hline
1 & 1 & 1 & 1 &-1 & -1 & 0 \\
\hline
1 & 1 & 1 & 1 & 0 & 0 & -1 \\
\hline
\multicolumn{6}{c}{B_u}
\end{array}  
&
\begin{array}{|rrrr|rr|r|}
\multicolumn{7}{r}{ }\\
\hline
\;\;0 &\;\; 0 &\;\; 1 &\;\;1 &\;\; 0 &\;\; 1 &\;\; 1 \\
0 & 1 & 0 & 1 & 0  & 0 & 1 \\
1 & 0 & 0 & 1 & 0 & 1 & 1 \\
1 & 1 & 1 & 0 & 1 & 1 & 1 \\
\hline
1 & 1 & 1 & 1 & 1 & 1 & 0 \\
\hline
1 & 1 & 1 & 1 & 0 & 0 & 1 \\
\hline
\multicolumn{6}{c}{B'_u}
\end{array}  
\end{array}
\]
\caption{Example of elementary matrix operations for D3$^+$.} \label{Fig:EMO-D3}
\end{figure}

\medskip
In the rest of this section, we discuss integer voting weights 
	and a quota of a rough voting representation. 
We say that a player $i \in N$ is 
	a {\em passer} if and only if 
	every coalition $S \ni i$ is winning. 			
	
\begin{theorem}
	Assume that a given simple game 
	$G=(N, {\cal W})$ satisfies 
	\mbox{$\emptyset \not \in {\cal W} \ni N.$} 
If a given simple game $G$ is roughly weighted,  
	then there exists an integer vector 
	$(q; \Vec{w}^{\top})$ of the rough voting representation 
	satisfying $0 \leq w_i \leq \alpha_{n-1} \;\; (\forall i \in N),$ 
	$0 \leq q \leq \alpha_n$, and 
	$1 \leq \sum_{i \in N}w_i \leq 2 \alpha_n.$
\end{theorem}

\noindent
Proof. 
First, we show that if a given game is roughly weighted,
	then either 
\[
	\mbox{P4:}
	\left(
	\begin{array}{rr}
		 A({\cal W}) & \Vec{0} \\
		-A({\cal L}) & \Vec{0} \\
		-\Vec{1}^{\top} & 1  
	\end{array}
	\right)
	\left(
	\begin{array}{c}
		\Vec{w} \\ u 
	\end{array}
	\right)
	\geq 
	\left(
	\begin{array}{r}
		\Vec{1}  \\
		-\Vec{1} \\
		0
	\end{array}
	\right), 
	\Vec{w} \geq \Vec{0}, u \geq 0,
\]	
	is feasible or there exists 
	at least one passer.
Suppose that a given simple game has a rough voting 
	representation $(q; \Vec{w}^{\top})$.
If  $q >0$, then  $ (1/q)\Vec{w} $
	becomes  a feasible solution of P4
	by setting $u$ to a sufficiently large positive number.
Consider the case $q \leq 0$.
Assumption $\emptyset \not \in  {\cal W}$
	implies that  $0 \leq q,$ and thus 
	we obtain  $q = 0$.
Properties $(q, \Vec{w}^{\top}) \neq \Vec{0}^{\top}$ 
	and $\Vec{w}\geq \Vec{0}$ 
	imply that  $\exists i^{\circ} \in N, w_{i^{\circ}} >0$, 
	i.e.,  a given game $G$ has a passer $i^{\circ}$.
		
When a given game $G$ has a passer $i^{\circ} \in N$, 
	then there exists a rough voting representation 
	$(q^{\circ}; \Vec{w^{\circ}}^{\top})$ defined by
\[
	w^{\circ}_i=\left\{ \begin{array}{ll}
			1 & (i=i^{\circ}), \\
			0 & (i \neq i^{\circ}), 
		\end{array}\right. 
		q^{\circ}=0, 
\]
which produces the desired result.

Lastly, we consider the case in which P4 is feasible.
It is well-known that when P4 is feasible, 
	there exists a basic feasible solution.
Let $(\Vec{w}'^{\top}, u')^{\top}$ 
	be a basic feasible solution of P4 and
	$B$ be a corresponding basis matrix. 
It is easy to see that 
	$(1; \Vec{w}'^{\top})$ is a  rough voting representation of $G$.
Assumption $N \in {\cal W}$ implies the positivity of $u'$
	because $u' \geq  \Vec{1}^{\top}\Vec{w}' \geq 1$.
Then, variable $u$ is a basic variable, and thus
	$B$ includes a column corresponding to $u$, 
	which is called the last column.
The non-singularity of $B$ implies that
	a column corresponding to $u$ is not the zero vector, 
	and thus $B$ includes a row corresponding to 
	the inequality $\Vec{1}^{\top}\Vec{w}\leq u$,
	which is called the last row
	(see Figure~\ref{Fig:EMOP4}).
The number of rows (columns) of 
	basis matrix $B$, denoted by $n'$,
	is less than or equal to $n+1$.

Cramer's rule states that  
	$(q^*, \Vec{w^*}^{\top}, u^*) = |\deT (B)| (1, \Vec{w}'^{\top}, u')$
	is a non-negative integer vector.
It is easy to see that 
	$(q^*, \Vec{w^*}^{\top}, u^*)$
	satisfies
\begin{eqnarray*}
	A({\cal W)}\Vec{w^*}&=&|\deT (B)| A({\cal W})\Vec{w}'
		\geq |\deT (B)| \Vec{1} =q^*\Vec{1}, \\
	A({\cal L})\Vec{w^*}&=&|\deT (B)| A({\cal L})\Vec{w}'
		\leq |\deT (B)| \Vec{1} =q^* \Vec{1}, \;\; \mbox{and} \\
	\Vec{1}^{\top}\Vec{w^*}&=& |\deT (B)| \Vec{1}^{\top}\Vec{w}' 
		\leq |\deT (B)|u'=u^*.
\end{eqnarray*}
From the above, $(q^*; \Vec{w^*}^{\top})$ is 
	an integer vector of a rough voting representation.
Assumption $N \in {\cal W}$ implies that
	$\Vec{1}^{\top}\Vec{w^*}\geq q^*=|\deT (B)|\geq 1$.
 
Let $\Vec{d}'_B$ be a  subvector of 
	the right-hand-side vector of an inequality system in P4
	corresponding to rows of $B$. 
Cramer's rule states that  
	$\det (B) u' =\deT (B_u),$ 
	where $B_u$ is obtained from $B$
	but the column corresponding to 
	a basic variable $u$
	is replaced by $\Vec{d}'_B$ 
	(see Figure~\ref{Fig:EMOP4}).
We multiply rows of  $B_u$ 
	 that correspond to losing coalitions by $(-1)$
	and multiply the last row by $(-1)$.
The resulting matrix, denoted by $B'_u$, 
	is a 0--1 matrix whose 
	last row includes exactly one 0-component
	(indexed by $u$).
Lemma~\ref{all-one}~(c2) implies that 
	$|\deT (B'_u)| \leq 2\alpha_{n'-1}
		\leq 2 \alpha_n$.
Thus, we obtain that 
\[
	\Vec{1}^{\top}\Vec{w^*} \leq u^* \leq |u^*|
	= | \deT (B) u' |
	= |\deT (B_u)| =|\deT (B'_u)|  \leq 2 \alpha_n.
\]
By analogy with  the proof 
	of Theorem~\ref{theorem-weightedness}, 
	we can prove the desired inequalities: 
	$q^*=|\deT (B)| \leq \alpha_n$ 
	and $w^*_i \leq \alpha_{n-1}$ 
	$(\forall i \in N)$.	
\QED

\begin{figure}[htb]
{\scriptsize 
\[
\begin{array}{|rrrrr|r|}
\multicolumn{1}{r}{w_1} & w_2 & w_3 & w_4 &
\multicolumn{1}{r}{w_5} &
\multicolumn{1}{r}{\;\;u} \\
\hline
1 & 1 & 1 & 0 & 1 & 0 \\
0 & 1 & 0 & 1 & 1 & 0 \\
\hline
0 &-1&-1 & 0 & 0 & 0 \\
-1& 0 & 0 &-1 &-1 &  0 \\
0 &-1 & 0 &-1 &  0 &  0 \\
\hline
-1&-1 &-1 &-1 &-1 & 1 \\
\hline
\multicolumn{5}{c}{B}
\end{array}
\quad
\begin{array}{|rrrrr|r|}
\multicolumn{1}{r}{w_1} & w_2 & w_3 & w_4 &
\multicolumn{1}{r}{w_5} &
\multicolumn{1}{r}{\;\;u} \\
\hline
1 & 1 & 1 & 0 & 1 & 1 \\
0 & 1 & 0 & 1 & 1 & 1 \\
\hline
0 &-1&-1 & 0 & 0 &-1 \\
-1& 0 & 0 &-1 &-1 &-1 \\
0 &-1 & 0 &-1 &  0 &-1 \\
\hline
-1&-1 &-1 &-1 &-1 & 0 \\
\hline
\multicolumn{5}{c}{B_u}
\end{array}
\quad
\begin{array}{|rrrrr|r|}
\multicolumn{1}{r}{w_1} & w_2 & w_3 & w_4 &
\multicolumn{1}{r}{w_5} &
\multicolumn{1}{r}{\;\;u} \\
\hline
1 & 1 & 1 & 0 & 1 & 1 \\
0 & 1 & 0 & 1 & 1 & 1 \\
\hline
0 & 1& 1 & 0 & 0 & 1 \\
1 & 0 & 0 & 1 & 1 & 1 \\
0 & 1 & 0 & 1 &  0 & 1 \\
\hline
 1& 1 & 1 & 1 & 1 & 0 \\
\hline
\multicolumn{5}{c}{B'_u}
\end{array}
\]
}
\caption{Examples of elementary matrix operations for P4.} \label{Fig:EMOP4}
\end{figure}

\section{Conclusion}

In this paper, 
we discussed the smallest value of $k^*$
	such that every $k^*$-trade robust simple game
	would be weighted.
We provided a new proof of the existence 
	of a trading transform 
	when a given simple game is non-weighted. 
Our proof yields an improved upper bound 
	on the required length of a trading transform.
We showed that a given simple game $G$ is weighted
	if and only if 
	$G$ is $\alpha_{n+1}$-trade robust, 
	where $\alpha_{n+1}$ denotes 
	the maximal value of determinants of 
	$(n+1) \times (n+1)$ 0--1 matrices.	
Applying the Hadamard's 
	evaluation~\citep{hadamard1893resolution} 
	of the determinant, we obtain 
	$ k^* \leq
	 \alpha_{n+1} \leq
	  (n+2)^{\frac{n+2}{2}}(1/2)^{(n+1)}$,
	which improves the existing bound 
	$ k^* \leq 
	 (n+1)n^{n/2}$ obtained by~\cite{gvozdeva2011weighted}.

Next, we discussed upper bounds 
	for the maximum possible integer weights and the quota 
	needed to represent any weighted simple game with $n$ players. 
We show that every weighted simple game 
	(satisfying $\emptyset \not \in {\cal W} \ni N$)
	has an integer-weight representation 
	$(q; \Vec{w}^{\top}) \in \Int \times \Int^N$ 
	such that
	$|w_i| \leq \alpha_n$ $(\forall i \in N)$,
	$|q| \leq \alpha_{n+1}$, and 
	$1 \leq \sum_{i \in N} w_i \leq 2 \alpha_{n+1}-1$.
We demonstrated the tightness of our bound on the quota
	when $n \leq 5$.

We described a rounding method 
	based on a linear relaxation of an integer programming problem
	for finding an integer-weight representation.
We showed that
	an integer-weight representation is obtained 
	by carefully rounding a solution of the linear inequality system
	multiplied by  $\lambda^{\bullet} \leq (2-\sqrt{2})n+(\sqrt{2}-1)	
	< 0.5858n+0.4143$.
Our proof of Theorem~\ref{bullet}  indicates an existence 
	of a randomized rounding algorithm for finding
	an appropriate value $\lambda^{\bullet}$.
However, from theoretical point of view, 
	Theorem~\ref{bullet} only showed the existence of 
	a real number $\lambda^{\bullet}$. 
Even if there exists an appropriate ``rational'' number $\lambda^{\bullet}$, 
	 we need to determine the size of the rational number 
	 (its numerator and denominator)
	 to implement a naive randomized rounding algorithm. 
Thus, it remains open whether there exists 
	an efficient algorithm for finding an integer-weight representation
	satisfying the properties in Theorem~\ref{bullet}. 

	
Lastly, we showed that  a roughly weighted simple game
	(satisfying $\emptyset \not \in {\cal W} \ni N$)
	has an integer vector 
	$(q; \Vec{w}^{\top})$ of the rough voting representation 
	satisfying 
	$0 \leq w_i \leq \alpha_{n-1} \;\; (\forall i \in N)$, 
	$0 \leq q \leq \alpha_n$,
	and $1 \leq \sum_{i \in N}w_i \leq 2 \alpha_n$.
When a given simple game is not roughly weighted, 
	we showed that
	(under the the monotonicity
	 property~(\ref{monotonicity}) and 
	  $\emptyset \not \in {\cal W} \ni N$) 
	there existed 
	a potent certificate of non-weightedness 
	whose length is less than or equal to 	
	$ 2 \alpha_{n+1}$.

\bibliographystyle{apalike}  

\bibliography{TradingTransformRefer.bib}   

\end{document}